\documentclass[prb,twocolumn,floatfix,footinbib,superscriptaddress,showpacs, showkeys]{revtex4-1}
\usepackage{graphicx}
\usepackage{amssymb}
\usepackage[utf8]{inputenc}
\usepackage[english]{babel}
\usepackage[usenames,dvipsnames]{color}
\usepackage{amsmath}

\newcommand{\vc}[1]{\ensuremath{\mathbf{#1}}}

\newcommand{\doubleindex}[2]{\tiny\begin{array}{l}{#1}\\{#2}\end{array}}

\begin{document}
\title{Electron energy spectrum of the spin-liquid state in a frustrated Hubbard model}
\author{A.~E.~Antipov}\email{antipov@ct-qmc.org}
\author{A.~N.~Rubtsov}
\affiliation{Department of Physics, Moscow State University,
119992 Moscow, Russia}

\author{M.~I.~Katsnelson}
\affiliation{Radboud University Nijmegen, Institute for Molecules
and Materials, 6525AJ Nijmegen, The Netherlands}

\author{A.~I.~Lichtenstein}
\affiliation{I. Institute of theoretical Physics, University of Hamburg, 20355
Hamburg, Germany}
\date{\today}

\begin{abstract}
Non-local correlation effects in the half-filled Hubbard model
on an isotropic triangular lattice are studied within a
spin polarized extension of the dual fermion approach. A competition between the antiferromagnetic non-collinear and the
spin liquid states is strongly enhanced by an incorporation of a k-dependent
self-energy beyond the local dynamical mean-field theory. The dual fermion corrections drastically decrease the energy of a spin liquid
state while leaving the non-collinear magnetic states almost non-affected. This makes the spin liquid to become a preferable state in a certain interval of interaction strength of an order of the magnitude of a bandwidth. The spectral function of the spin-liquid Mott insulator is determined by a formation of local
singlets which results in the energy gap of about twice larger than that of the 120 degrees antiferromagnetic Neel state.
\end{abstract}


\maketitle

\section{Introduction}
In some special cases, quantum fluctuations can destroy an ordered state
even at zero temperature. The study of such ``quantum phase
transitions'' due to zero-point fluctuations is a hot subject in
modern condensed matter physics \cite{Sachdev:Book}. These
fluctuations are especially important in frustrated systems. A
prototype of classical frustrated spin system is the Ising model on a 
triangular lattice with nearest-neighbor antiferromagnetic
interaction. The ground state of this model has a macroscopically
large degeneracy and the addition of quantum spin-flip terms results into
a formation of spin-liquid state known as ``resonating valence
bonds'' (RVB) \cite{Anderson:MatResBull:1973,
Fazekas:PhilMag:1974}.  There are strong experimental evidences that the RVB is the ground state
of some \textit{Mott insulators:} organic salt
$\kappa$-(ET)$_2$Cu$_2$(CN)$_3$\cite{Shimizu:PRL:2003,Yamashita:NaturePhysics:2008,
Helton:PRL:2007} and spinel oxide on the frustrated
hyperkagome lattice Na$_4$Ir$_3$O$_8$\cite{Okamoto:PRL:2007}. 

Theoretical description of the materials with the RVB Mott insulating
state requires an accurate choice of parameters of the model. The Hubbard model on a triangular lattice with nearest-neighbor
hopping is the simplest choice, though its validity can be questioned. First, recent \textit{ab-initio}
calculations\cite{Kandpal:PRL:2009,Nakamura:JPSJ:2009} suggest
that the family of $\kappa-(CN)$ Mott insulating organic salts has
a large anisotropy of approximately $20\%$ difference between hopping parameters in different directions.
Moreover, it is not clear which phase of the Hubbard model on a isotropic triangular lattice
is a ground state. The RVB phase and 120 degrees antiferromagnetic Neel
states are expected to possess very close energies\cite{Yoshioka:PRL:2009,Imada:Comment:2008}. Any of these phases can be stabilized in the
experiment by tuning next-neighbor hopping, electron-phonon interaction, anisotropy etc. This makes the results for the ground state obtained within the
Hubbard model on an isotropic triangular lattice not straightforwardly comparable with an experiment.

In the limit of a strong on-site repulsion $U$ this model is equivalent to the Heisenberg model
with antiferromagnetic nearest-neighbor exchange interaction $J =
2t^2/U$ where $t$ is the hopping parameter. The classical
Heisenberg model predicts a noncollinear magnetic structure with
$120^{\circ}$ angles between neighboring spins
\cite{Katsura:JSOP:1986}. Properties of quantum Heisenberg model
are likely to be similar and several theoretical and numerical QMC and DMRG
results predict the Neel $120^{\circ}$ ordering of ground
state\cite{Jolicoeur:PRB:1990, Bernu:PRB:1994, Capriotti:PRL:1999,
White:PRL:2007}. It was suggested that a disordered state of quantum Heisenberg antiferromagnet may behave as a spin liquid\cite{Kalmeyer:PRB:1989}.

For moderate values of $U$, the situation is less clear. The
results obtained by a \textit{path integral renormalization group}
(PIRG)\cite{Morita:JPSC:2002,Yoshioka:PRL:2009} and the
\textit{variational cluster approach}
(VCA)\cite{Sahebsara:PRL:2008} suggest the non-magnetic insulating
state to prevail at zero temperatures in certain interval of
parameter $U/t$ and the Neel ordered state to exist at higher
values of $U$.

To study the Hubbard model for the strongly correlated
lattice systems, the dynamical mean-field theory (DMFT)
\cite{MetznerVollhardt:PRL:1989, Georges:RevModPhys:1996, Kotliar:RevModPhys:2006} is frequently used.
This approach has been employed to the case of triangular
lattice \cite{Lu:arxiv:2007, Aryanpour:PRB:2006}, however, the {\it local} DMFT scheme is not sufficient to study the RVB physics
based on the formation of singlets on bonds \cite{Anderson:MatResBull:1973, Fazekas:PhilMag:1974,Anderson:Book:1997} which
involve non-locality, at least, at the level of pairs of sites.
Recently, non-local methods beyond the DMFT such as the cluster extension of DMFT \cite{Licht:PRB:2000}, dynamical
cluster approximation (DCA) \cite{Jarrell:RevModPhys:2005, Hettler:PRB:1998} and dual fermions (DF)
theory \cite{Rubtsov:PRB:2008,Rubtsov:PRB:2009} have been applied to the problem \cite{Liebsch:PRB:2009, LeeMonien:PRB:2008, Kyung:PRB:2007}.
However, the competition of the two phases has not been considered there
and their electron spectral function phases has not been discussed. 

In this paper a spin-polarized extension of the DF technique is
implemented for the isotropic triangular lattice. Contrary to the local DMFT, the
long-ranged electron correlations are taken into account within a regular diagrammatic perturbation scheme
starting from the DMFT as zero-order approximation \cite{Rubtsov:PRB:2008}. It is shown that these corrections change
essentially the phase diagram of a triangular lattice decreasing the energy of the spin-liquid
state in comparison with the DMFT results. For the 120$^{\circ}$ Neel State the corrections
are much smaller so that the spin liquid state has a lower total energy than the Neel-ordered phase in a certain range of parameters. We calculate the electron energy spectrum by means of the DF scheme utilizing the
continuous time quantum Monte Carlo solver \cite{Rubtsov:PRB:2005} at low
temperatures.

\section{Method}

In the spirit of a mean-field theory the whole lattice problem is mapped to the single impurity, in this case to a single-impurity Anderson model subject to a self-consistent condition. Such a dynamical mean-field theory (DMFT) restricts an investigation to studying only local correlation effects. For frustrated quantum systems this approximation is obviously not applicable. In order to take into account non-local corrections two approaches are implemented: (i) the long-ranged correlations for the anti-ferromagnetic ordering are included by introducing the spin polarization to the effective bath, (ii) the short-ranged correlations are incorporated within the dual fermion technique - it is based on a transformation to new dual fermionic variables, which make the diagrammatic expansion over non-linearity much more effective\cite{Rubtsov:PRB:2008,Rubtsov:PRB:2009}.

We start with a single-site DMFT-like problem. For a given impurity with a yet undefined dynamical hybridization function $\Delta_{\omega\sigma}$, which describes the interaction with a fermionic bath of surrounding lattice, the effective action for single-impurity Anderson model is written in the following way:
\begin{equation}
S_{imp}=\sum_{\omega\sigma} (\Delta_{\omega\sigma}-\mu-i \omega)
c^\dagger_{\omega\sigma} c_{\omega\sigma} + U\int_0^\beta n_{\uparrow
\tau} n_{\downarrow \tau}  d\tau,
\end{equation}
where index $\sigma=\uparrow,\downarrow$ labels the spin projections, $U$ is an a on-site Coulomb interaction, $\mu$ is a chemical potential, and $\beta$ is an  inverse temperature. The solution of this problem can be obtained numerically by the CT-QMC solver\cite{Rubtsov:PRB:2005}. Along with the one-particle Green's function $g_{\omega\sigma}=-i\langle c_{\omega\sigma}c^{\dagger}_{\omega\sigma}\rangle$ one can obtain the higher order correlators, such as the two-particle Green's function
$\chi_{1234}=\langle{\mathbb T}
c_{\sigma_1}(\tau_1)c_{\sigma_2}(\tau_2)c^{\dagger}_{\sigma_3}(\tau_3)c^{\dagger}_{\sigma_4}(\tau_4)\rangle
$ and the corresponding vertex function
$$\gamma^{(4)}_{1234} = g_{11'}^{-1}g_{22'}^{-1} \left[
\chi_{1'2'3'4'} - (g_{1'3'}g_{2'4'} - g_{1'4'}g_{2'3'}) \right]
g_{3'3}^{-1}g_{4'4}^{-1}$$
We now need to establish the connection
between the single-site model and original lattice problem. The whole lattice is divided into
supercells, combined of three atoms. The Fourier transforms over both the supercell and cluster spatial indices are utilized, which corresponds to the DCA-like scheme\cite{Jarrell:RevModPhys:2005}:
\begin{multline}G(\vc r =\vc R_M +  \vc{r_i} , \vc r' = \vc R_{M'} +  \vc{r_j}) \\=\sum_{\doubleindex{\vc K}{\mathbf{k}\mathbf{k'}}} G_{\mathbf{k}\mathbf{k'}}(\vc K)e^{i\vc K (\vc R_m - \vc R_{M'})}e^{i(\mathbf{k}\vc r_i - \mathbf{k'}\vc{r_j})},
\end{multline}
$\vc R_M$ is a radius vector of the supercell, $\vc r_i$ points to a site inside it. $\vc K$ is defined inside a Brillouin zone of a supercell, \vc{k} is its basis vector. In this notation, the Hamiltonian of the Hubbard model can be written as follows:
\begin{equation}
\hat H = \sum_{\vc{K}\vc{k}\vc{k'}\sigma} c^{\dagger}_{\vc{k}\vc K\sigma}(\varepsilon_{\vc{k}\vc{k'}}(\vc K)-\mu)c_{\vc{k'}\vc K\sigma} + \sum_i U \hat n_{i\uparrow}\hat n_{i\downarrow},
\end{equation}
where index $i$ labels a of site of the lattice. For the supercell approach the dispersion law is given by $\varepsilon_{\vc{k}\vc{k'}}(\vc K)=\varepsilon(\vc K + \vc{k})\delta_{\vc{k}\vc{k'}}$ with the spectrum of a triangular lattice:
\begin{multline}\varepsilon(\vc K = \alpha\vc{k_1}+\beta\vc{k_2}) =
-2t\Bigl[\cos(\frac{2\pi}{3}(-\alpha+\beta))+ \\
+\cos(\frac{2\pi}{3}(\alpha+2\beta)) + \cos(\frac{2\pi}{3}(-2\alpha-\beta))
\Bigr],  \end{multline}
where $\vc{k_1},\vc{k_2}$ are non-collinear basis vectors of the supercell in the reciprocal space: $\vc{k_1}=\frac{1}{3}(\vc{k_x}+\vc{k_y})$, $\vc{k_2}=\frac{1}{3}(-\vc{k_x}+2\vc{k_y}), |\vc{k_1}| = |\vc{k_2}|$, where $\vc k_x, \vc k_y$ are the basis vectors of the reciprocal triangular lattice.

After adding and subtracting an arbitrary hybridization function into the action for the Hubbard model, one can obtain:
\begin{equation}
S[c,c^*] = \sum_i S_{imp}^{(i)} - \sum_{\vc K\vc{k}\vc{k'}\omega\sigma}\vc{c}^*_{\vc K\vc k\sigma}(\Delta_{\omega\vc{k}\vc{k'}} - \varepsilon_{\vc{k}\vc{k'}}(\vc K))\vc{c}_{\vc K\vc k'\sigma}
\end{equation}
It should be mentioned that the $\vc k$ dependence of $\Delta_\omega$ arises from the spatial Fourier transform over internal sites of the supercell. In case of all sites being equivalent $\Delta_{\omega\vc{k}\vc{k'}} = \Delta_\omega\delta_{\vc{k}\vc{k'}}$.

The further steps of the dual fermion scheme for a supercell are similar to the cluster DF-approach \cite{Hafermann:2007:JETPLett}. By utilizing Hubbard-Stratonovich transformation to new fermionic variables and integrating out the original degrees of freedom one can obtain a new fermionic problem, which is even more complex that the original one, but it allows to be efficiently treated by a diagrammatic expansion by a right definition of a yet arbitrary hybridization function $\Delta_{\omega\sigma}$. It is chosen in a way for the trivial case of no diagrammatic expansion of the self energy of the dual variables to be equivalent to the DMFT. We use a first non-zero term in the diagrammatic expansion of a dual self-energy:
\begin{multline} \label{dual_correction}
 \Sigma^{\text{d}}_{11'}=-\frac{1}{2\beta^2}\sum_{
\doubleindex{234}{2'3'4'}}
\delta_{\omega_1+\omega_3}^{\omega_2+\omega_4}
\gamma^{(4)}_{1234}\\G^{\text{d}}_{44'}
(\vc R)G^{\text{d}}_{2'2}(-\vc R)G^{\text{d}}_{3'3}(\vc R)\gamma^
{(4)}_{4'3'2'1'}
\end{multline}
The important long ranged corrections to the single-site DMFT scheme are introduced in formula (\ref{dual_correction}). It should be noted that the dual fermion Green's function consists only of non-local part while the vertex function is local and obtained within a single-impurity Anderson model\cite{Rubtsov:PRB:2008,Rubtsov:PRB:2009}. 

In order to introduce non-collinear spin-dependent DF-scheme one can rotate spinors for different sites of a supercell. The local Green's function for a given site is obtained as follows:
\begin{equation} g_{\omega i}=\hat U_i(\frac{\pi}{2},\pm\frac{2\pi}{3}) \hat g_\omega \hat U_i^*(\frac{\pi}{2},\pm\frac{2\pi}{3}), \end{equation}
where $\hat U$ is a spin $2\times2$ rotation matrix.

The total energy $\langle \hat H - \mu \hat N \rangle$ for different states has been calculated in the following form \cite{ Kotliar:RevModPhys:2006}:
\begin{eqnarray}
\langle \hat H \rangle = \langle \hat H_0 +\hat  V - \mu \hat N \rangle, \\
\langle H_0 \rangle = -t \sum_{j\sigma} G_{0j} (\tau=0+), \\
\langle H_0 + 2 V - \mu N \rangle = \left(\frac{\partial G_{00}(\tau)}{\partial \tau}\right)_{\tau \rightarrow 0+}
\end{eqnarray}
This procedure requires the Fourier transform of the Green's function to the
imaginary time domain. Since the CT-QMC solver provides a bare Green's function on
finite number of Matsubara frequencies, this could be a serious
source of numerical errors. The interaction expansion solver \cite{Rubtsov:PRB:2005} produces
low noise at the high Matsubara frequencies and allows to
fit an asymptotic tail of the Green's Function.
\section{Results}
In order to calculate the energetics of a triangular lattice, 2-3 different independent runs consisting of 20-30 DMFT iterations and following 15-25 DF iterations were done with the lattice size of $32\times32$. The interaction expansion QMC solver \cite{Rubtsov:PRB:2005} was used for which the QMC error becomes sufficient at high values of $U$. The obtained data was averaged, and the corresponding statistical error is plotted on the graphs. The spin-polarized phase in the insulating regime hence being constructed by spin rotation represents the Neel state. The unpolarized DMFT insulating phase has only local correlations and does not possess any magnetic ordering. It will be shown that the non-polarized insulating DF phase is constructed by singlet corrections to a local self-energy and therefore perturbatively describes the spin-liquid state. 
\subsection{Total energy calculations}
\begin{figure}[t]\label{1}
\includegraphics[width=1.0\columnwidth]{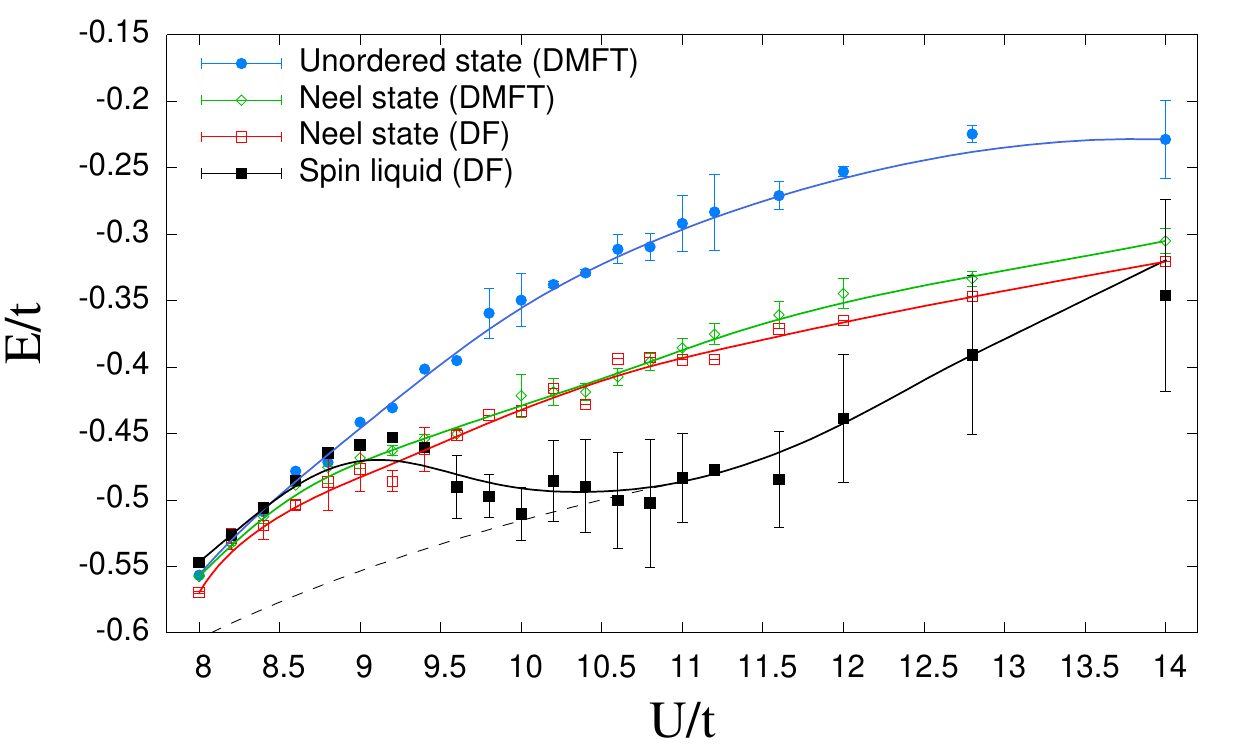}
\caption{(Color online) The dependence of the total energy per site at $\beta t=10$
on the parameter $U$. The separate curves for a phase transition from metallic to a spin liquid and from metallic to a Neel state were obtained. The spin liquid state possesses lower energy in the interval $9.5 < U/t < 13$. At higher values of $U$ the energies of these insulating states become close. The phase transition from metallic to the spin liquid phase occurs at $U/t \approx 9.6 \pm 0.2$, the value of the critical $U/t$ for the Mott transition from metallic to the 120$^\circ$ Neel phase lies in the interval $8.25 - 9.4 \pm 0.2$. The dashed line represents the monotonic interpolation of the spin liquid energy curve. }
\end{figure}
\begin{figure}[t]
\includegraphics[width=1.0\columnwidth]{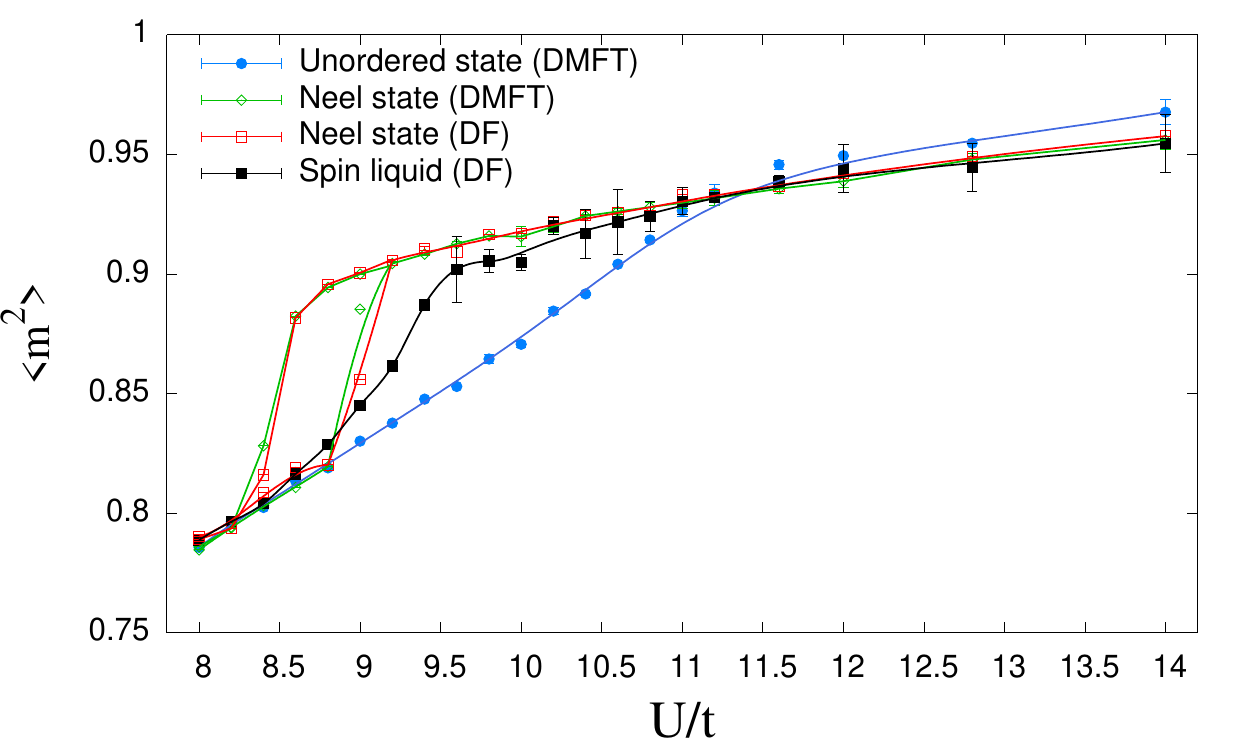}
\caption{(Color online) The dependence of the averaged square of magnetic moment per site at $\beta t=10$ on the parameter $U$. With the increasing of $U$ the formation of magnetic state is detected first for the 120$^0$ Neel state at $U/t = 8.25 - 9.4 \pm 0.2$. The hysteresis in the Neel curve represents a first-order phase transition from metallic to a Neel state. The formation of the magnetic moments for the spin liquid state occurs at the value of critical $U$ of a Mott phase transition $U=9.6 \pm 0.2$. }
\end{figure}
The results for a total energy per site $\langle H+\mu N \rangle = \langle H \rangle + U/2$ and averaged square of magnetic moment $\langle m^2 \rangle$ as a function of $U/t$ in the region of $8<U/t<14$ for various states at half-filling, $\beta t = 10$ are presented at Figures 1 and 2 respectively. The separate curves for the phase transition from a metallic to a spin liquid state and from a metallic to a 120$^0$ Neel state were obtained. The phase transition to a spin liquid state occurs at $U/t = 9.6 \pm 0.2$, the hysteresis at $U/t=8.25 - 9 \pm 0.2$ represents a first order transition from the metallic to the Neel ordered state. 

At first, it can be seen from the energy graph that there is an essential difference between the DMFT and DF results for the spin liquid case. Non-local corrections favour the unordered insulating state. Without DF-corrections this phase never corresponds to the ground state in the whole studied region and remains metallic up to $U/t<11.2 \pm 0.2$. This is in agreement with the previous DMFT calculations\cite{Aryanpour:PRB:2006}. Thus the first non-vanishing diagram in the DF approach lowers the total energy of the unordered insulating state to make it favourable in the region of $9.5<U/t<13$. The dual corrections in the Neel phase are small and do not change the behaviour of this phase. At higher values of $U/t$ the curves of total energies of Neel and spin liquid state become close and cannot be distinguished within the statistical error of the method. Nevertheless, since the dual corrections vanish at high $U$ \cite{Rubtsov:PRB:2008} one would expect that the total energies of DMFT and DF curve will become equal, with the Neel ordering curve to dominate in this region. 

A formation of local magnetic moments with the increasing of $U/t$ is captured in the Figure 2 ($\beta t = 10$). It can be seen that in the insulating region both spin liquid and $120^0$ Neel states possess well-defined magnetic moments thus differing by a type of ordering. The dual correction for the unordered state significantly enhances a formation of magnetism in the region of Mott transition, thus emphasizing the role of singlet correlations. As for the Neel phase these corrections are small and magnetic ordering is formed even at lower value of $U$, which means that a static spin ordering is essentially captured on the DMFT level. 
\subsection{The density of states and the nonlocal self-energy}
The densities of states $N(E)$ for the strongly correlated metallic
state are shown in the Figure 3. One can
clearly see a three-peak structure typical for the systems at the
metal side of Mott transition, namely, two Hubbard bands and
central ``Kondo'' quasiparticle peak \cite{Kotliar:RevModPhys:2006}. This
general structure is very similar both in the local DMFT and the non-local DF
approximations. There is an essential difference between two schemes related to the shape
of the Kondo peak. The density of states at the Fermi level in
the local approximation is fixed by the sum rule, namely, $N(0)$
should be equal to its value for noninteracting electrons
\cite{Kotliar:RevModPhys:2006}. In the DF approach, this is no more the case and
one can see that the non-local corrections suppress the height of
the Kondo peak at larger $U$.

\begin{figure}[h]
\includegraphics[width=1.0\columnwidth]{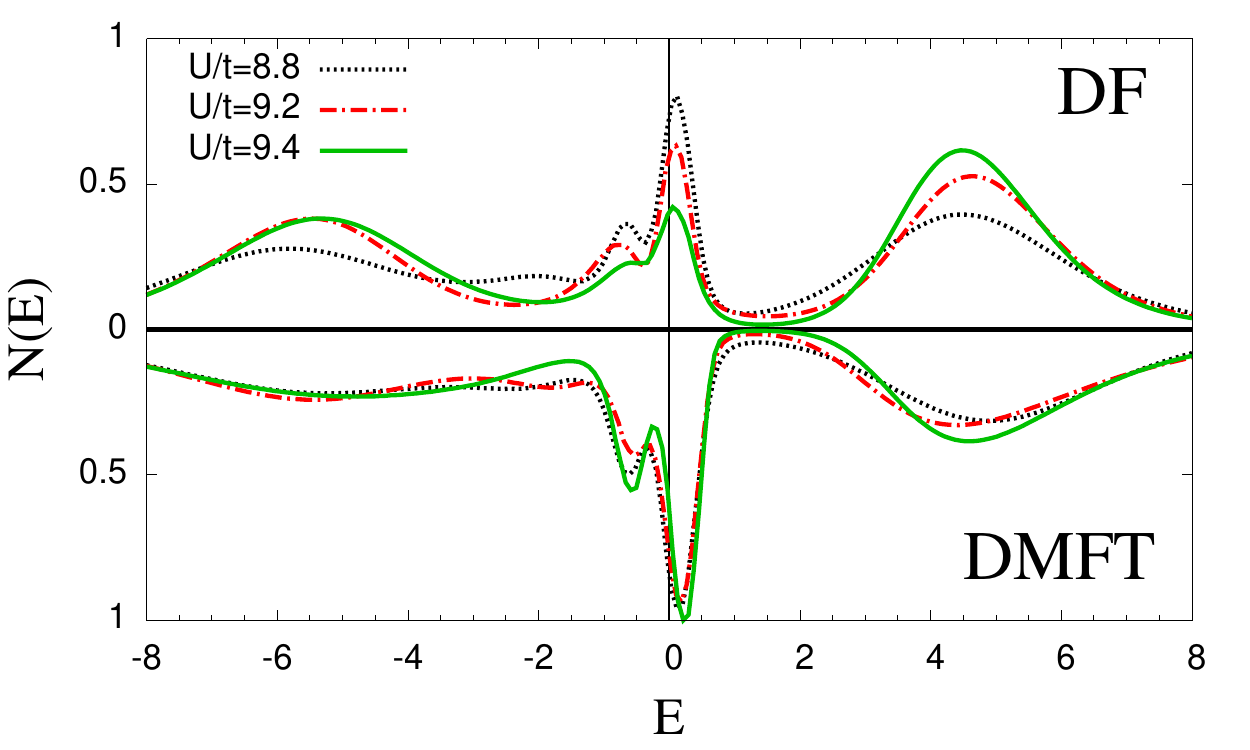}
\caption{(Color online) The density of states $N(E)$ of a metallic phase at $\beta t=10$ and $U/t=8.8, 9.2, 9.4$. The upper part of the graph corresponds to a dual fermion solution, the lower part is obtained by the DMFT. 
The height of the central ``Kondo" peak calculated by a DF technique is not fixed.}
\end{figure}
\begin{figure}[ht]
\includegraphics[width=1.0\columnwidth]{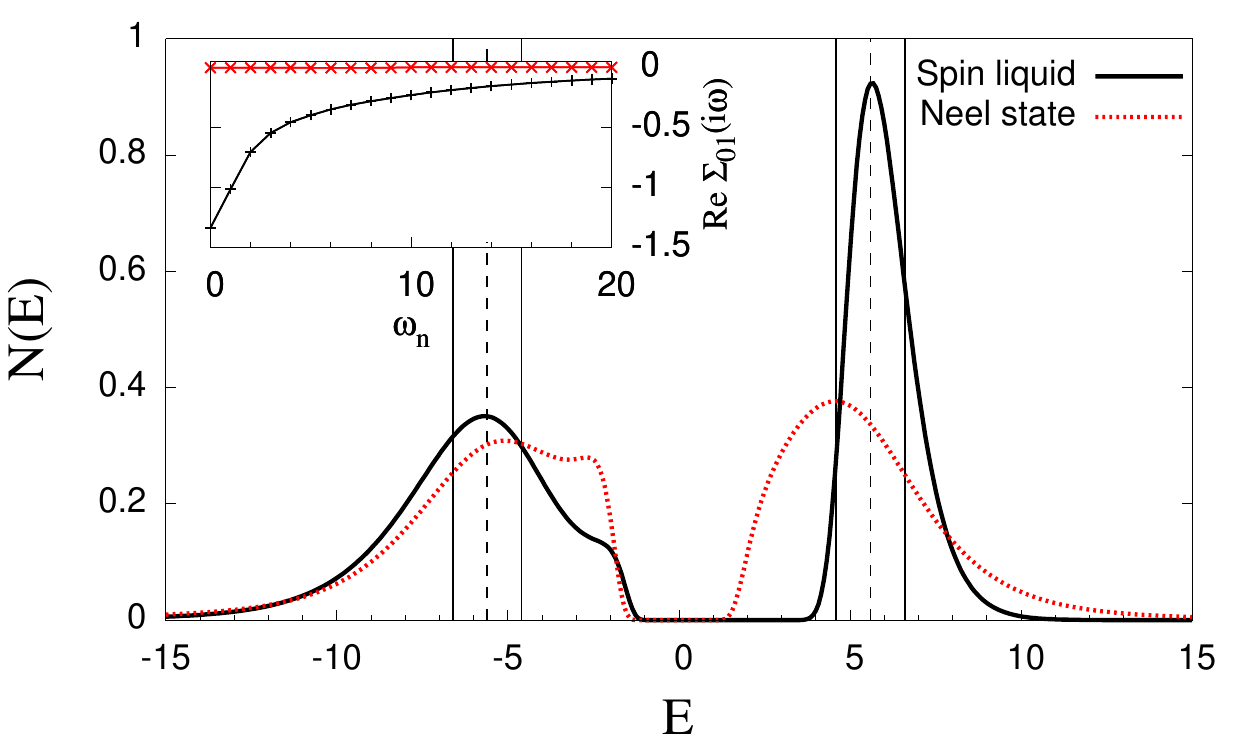}
\caption{(Color online) The density of states $N(E)$ of
the insulator phase at $\beta t=15$ and $U/t=11.2$ for
Neel and spin liquid states. The
difference between insulating gaps is seen. The spectrum of the 2-atom singlet with the same values of $U$ and $t$ is plotted by black ticks. The grey ticks represent the value $\pm U/2$. The inset shows the
real part of the nearest-neighbor self-energy over Matsubara
frequencies in spin liquid regime (black). The same value
obtained for Neel state (red) is negligible, while non-local
self-energy function in DMFT is by definition equal to zero.}
\end{figure}
The densities of states for the Neel and spin liquid states are
shown in the Figure 4. One can see that the spin liquid
solution has a wider gap than the Neel state. This agrees with the RVB-picture as a liquid formed by local singlets \cite{Anderson:MatResBull:1973}. Indeed,
local singlet must have a spectrum inherited from a spectrum of a
two-site cluster. To emphasize this feature, we show a spectrum of the isolated 2-atom cluster with the same value of $t$ with black ticks in the Fig. 4. The grey ticks represent value $\pm U/2$. As one can see from the inset of the Fig. 4, the non-local
corrections to nearest-neighbor self-energy in the spin-liquid case
are much larger than in the Neel state and dominate the
low-frequency spectrum. This reflects an importance of {\it
nonlocal dynamical fluctuations} for the spin liquid state. On the contrary, for
the Neel phase a most of the nonlocality is reflected in the {\it
static} ordering and is therefore taken into account already at the DMFT
level. This corresponds to the results of the Fig. 2.

\section{Discussion}
Some approximations have been made in our consideration, and they
should be discussed in the context of the phase diagram of the model
studied. The main of them are a finite-temperature consideration and taking into account just the leading dual correction.

The dual fermion corrections are, in general, sufficient in the
intermediate values of $U/t$ at the order of a magnitude of the bandwidth. The
picture obtained by including first non-zero diagram corresponds to an
idea of a spin-liquid as a state which can be observed in certain
region of parameter $U/t$ with metallic and Neel ordered states to
occur outside of this range. This corresponds to what is provided
in the literature\cite{Yoshioka:PRL:2009,
Sahebsara:PRL:2008,Kyung:PRL:2006}. At the Fig. 2 it was shown that an
inclusion of singlet corrections sufficiently enhances the magnetism
and makes the formation of magnetic state to occur at lower values of
$U$. Our calculations have been done using a leading diagram only and
we expect higher order diagrams to shift the value of $U_{c1}$ even
lower, so that a spin liquid state should have a monotonically
increasing total energy curve - that is illustrated by a dashed line in the Fig. 1.
We expect a spin liquid ground state to be extended to the region beyond current $U_{c1} = 9.6$ and a Neel ordered state to appear after the spin liquid. To have more quantitative description of a phase diagram one has to
include a series of diagrams, such as ladder, which may lead to a
formation of bounded singlet states\cite{Hafermann:PRL:2009}.

The upper critical value $U_{c2}\geq14$ looks overestimated due to a
finite temperature. Indeed, thermal fluctuations in general destroy an
ordering. The most clear case is a half-filled Hubbard lattice, where
an ordered ground state is realised at any (even infinitesimal) $U$, due
to so-called perfect nestling. However, DMFT calculations at
relatively small temperature $\beta=20$ show a phase diagram with the
ordering destroyed by local fluctuations of magnetic momenta up to
$U\approx 1$[\cite{Rubtsov:PRB:2009}].
Speaking about the energy dependencies presented in the Fig. 1, we
expect that lowering the temperature should shift the energy of the
Neel state down.

\section{Conclusions}
In summary, the Hubbard model on a triangular lattice is investigated within a framework of spin-polarized dual fermion approach. The ordering of 120${}^\circ$ Neel phase is described by an introduction of a non-collinear spin-polarization via a spinor rotation to the effective bath. The DMFT approximation works well for this type of ordering since non-local correlations are relatively small. On the contrary the solution with a non spin-polarized bath which describes the spin-liquid state has strong non-local corrections. This leads to a significant change in the phase diagram at low temperatures - the DF approach produces an insulating state with the lower total energy compared to that of a 120$^\circ$ state in a certain region of parameter $U/t$. Note that the simple DMFT solution gives an artificial Fermi liquid at these parameters. The densities of states for a spin-liquid phase has much larger insulating gap than for a non-collinear Neel state. Due to sufficient non-local corrections the dual fermion approach can capture the essential physics of a spin-liquid phase. 

Authors are grateful to H. Hafermann for providing the initial code for the DF calculations. This work was supported by DFG Grant No. 436 113/938/0-R and RFBR Grants No. 08-02-91953 and 08-03-00930-a. MIK acknowledges a financial support from the EU-India FP-7 collaboration under MONAMI.

\bibliography{heisenberg}

\begin{thebibliography}{10}%
\makeatletter
\providecommand \@ifxundefined [1]{%
 \ifx #1\undefined \expandafter \@firstoftwo
 \else \expandafter \@secondoftwo
\fi
}%
\providecommand \@ifnum [1]{%
 \ifnum #1\expandafter \@firstoftwo
 \else \expandafter \@secondoftwo
\fi
}%
\providecommand \enquote [1]{``#1''}%
\providecommand \bibnamefont  [1]{#1}%
\providecommand \bibfnamefont [1]{#1}%
\providecommand \citenamefont [1]{#1}%
\providecommand\href[0]{\@sanitize\@href}%
\providecommand\@href[1]{\endgroup\@@startlink{#1}\endgroup\@@href}%
\providecommand\@@href[1]{#1\@@endlink}%
\providecommand \@sanitize [0]{\begingroup\catcode`\&12\catcode`\#12\relax}%
\@ifxundefined \pdfoutput {\@firstoftwo}{%
 \@ifnum{\z@=\pdfoutput}{\@firstoftwo}{\@secondoftwo}%
}{%
 \providecommand\@@startlink[1]{\leavevmode\special{html:<a href="#1">}}%
 \providecommand\@@endlink[0]{\special{html:</a>}}%
}{%
 \providecommand\@@startlink[1]{%
  \leavevmode
  \pdfstartlink
   attr{/Border[0 0 1 ]/H/I/C[0 1 1]}%
   user{/Subtype/Link/A<</Type/Action/S/URI/URI(#1)>>}%
  \relax
 }%
 \providecommand\@@endlink[0]{\pdfendlink}%
}%
\providecommand \url  [0]{\begingroup\@sanitize \@url }%
\providecommand \@url [1]{\endgroup\@href {#1}{\urlprefix}}%
\providecommand \urlprefix [0]{URL }%
\providecommand \Eprint[0]{\href }%
\@ifxundefined \urlstyle {%
  \providecommand \doi [1]{doi:\discretionary{}{}{}#1}%
}{%
  \providecommand \doi [0]{doi:\discretionary{}{}{}\begingroup
  \urlstyle{rm}\Url }%
}%
\providecommand \doibase [0]{http://dx.doi.org/}%
\providecommand \Doi[1]{\href{\doibase#1}}%
\providecommand \bibAnnote [3]{%
  \BibitemShut{#1}%
  \begin{quotation}\noindent
    \textsc{Key:}\ #2\\\textsc{Annotation:}\ #3%
  \end{quotation}%
}%
\providecommand \bibAnnoteFile [2]{%
  \IfFileExists{#2}{\bibAnnote {#1} {#2} {\input{#2}}}{}%
}%
\providecommand \typeout [0]{\immediate \write \m@ne }%
\providecommand \selectlanguage [0]{\@gobble}%
\providecommand \bibinfo [0]{\@secondoftwo}%
\providecommand \bibfield [0]{\@secondoftwo}%
\providecommand \translation [1]{[#1]}%
\providecommand \BibitemOpen[0]{}%
\providecommand \bibitemStop [0]{}%
\providecommand \bibitemNoStop [0]{.\EOS\space}%
\providecommand \EOS [0]{\spacefactor3000\relax}%
\providecommand \BibitemShut [1]{\csname bibitem#1\endcsname}%
\bibitem{Sachdev:Book}%
  \BibitemOpen
  \bibfield{author}{%
  \bibinfo {author} {\bibfnamefont{S.}~\bibnamefont{Sachdev}},\ }%
  \emph{\bibinfo {title} {Quantum Phase Transitions}}\ (\bibinfo {publisher}
  {Cambridge Univ. Press},\ \bibinfo {year} {1999})%
  \bibAnnoteFile{NoStop}{Sachdev:Book}%
\bibitem{Anderson:MatResBull:1973}%
  \BibitemOpen
  \bibfield{author}{%
  \bibinfo {author} {\bibfnamefont{P.~W.}\ \bibnamefont{Anderson}},\ }%
  \bibfield{journal}{%
  \bibinfo {journal} {Materials Research Bulletin}\ }%
  \textbf{\bibinfo {volume} {8}},\ \bibinfo {pages} {153} (\bibinfo {year}
  {1973})%
  \bibAnnoteFile{NoStop}{Anderson:MatResBull:1973}%
\bibitem{Fazekas:PhilMag:1974}%
  \BibitemOpen
  \bibfield{author}{%
  \bibinfo {author} {\bibfnamefont{P.}~\bibnamefont{Fazekas}}\ and\ \bibinfo
  {author} {\bibfnamefont{P.~W.}\ \bibnamefont{Anderson}},\ }%
  \bibfield{journal}{%
  \Doi{10.1080/14786439808206568}{\bibinfo {journal} {Philosophical Magazine}}\
  }%
  \textbf{\bibinfo {volume} {30}},\ \bibinfo {pages} {423 } (\bibinfo {year}
  {1974})%
  \bibAnnoteFile{NoStop}{Fazekas:PhilMag:1974}%
\bibitem{Shimizu:PRL:2003}%
  \BibitemOpen
  \bibfield{author}{%
  \bibinfo {author} {\bibfnamefont{Y.}~\bibnamefont{Shimizu}}, \bibinfo
  {author} {\bibfnamefont{K.}~\bibnamefont{Miyagawa}}, \bibinfo {author}
  {\bibfnamefont{K.}~\bibnamefont{Kanoda}}, \bibinfo {author}
  {\bibfnamefont{M.}~\bibnamefont{Maesato}},\ and\ \bibinfo {author}
  {\bibfnamefont{G.}~\bibnamefont{Saito}},\ }%
  \bibfield{journal}{%
  \Doi{10.1103/PhysRevLett.91.107001}{\bibinfo {journal} {Phys. Rev. Lett.}}\
  }%
  \textbf{\bibinfo {volume} {91}},\ \bibinfo {pages} {107001} (\bibinfo {month}
  {Sep}\ \bibinfo {year} {2003})%
  \bibAnnoteFile{NoStop}{Shimizu:PRL:2003}%
\bibitem{Yamashita:NaturePhysics:2008}%
  \BibitemOpen
  \bibfield{author}{%
  \bibinfo {author} {\bibfnamefont{S.}~\bibnamefont{Yamashita}}, \bibinfo
  {author} {\bibfnamefont{Y.}~\bibnamefont{Nakazawa}}, \bibinfo {author}
  {\bibfnamefont{M.}~\bibnamefont{Oguni}}, \bibinfo {author}
  {\bibfnamefont{Y.}~\bibnamefont{Oshima}}, \bibinfo {author}
  {\bibfnamefont{H.}~\bibnamefont{Nojiri}}, \bibinfo {author}
  {\bibfnamefont{Y.}~\bibnamefont{Shimizu}}, \bibinfo {author}
  {\bibfnamefont{K.}~\bibnamefont{Miyagawa}},\ and\ \bibinfo {author}
  {\bibfnamefont{K.}~\bibnamefont{Kanoda}},\ }%
  \bibfield{journal}{%
  \bibinfo {journal} {Nat Phys}\ }%
  \textbf{\bibinfo {volume} {4}},\ \bibinfo {pages} {459} (\bibinfo {month}
  {06}\ \bibinfo {year} {2008})%
  \bibAnnoteFile{NoStop}{Yamashita:NaturePhysics:2008}%
\bibitem{Helton:PRL:2007}%
  \BibitemOpen
  \bibfield{author}{%
  \bibinfo {author} {\bibfnamefont{J.~S.}\ \bibnamefont{Helton}}, \bibinfo
  {author} {\bibfnamefont{K.}~\bibnamefont{Matan}}, \bibinfo {author}
  {\bibfnamefont{M.~P.}\ \bibnamefont{Shores}}, \bibinfo {author}
  {\bibfnamefont{E.~A.}\ \bibnamefont{Nytko}}, \bibinfo {author}
  {\bibfnamefont{B.~M.}\ \bibnamefont{Bartlett}}, \bibinfo {author}
  {\bibfnamefont{Y.}~\bibnamefont{Yoshida}}, \bibinfo {author}
  {\bibfnamefont{Y.}~\bibnamefont{Takano}}, \bibinfo {author}
  {\bibfnamefont{A.}~\bibnamefont{Suslov}}, \bibinfo {author}
  {\bibfnamefont{Y.}~\bibnamefont{Qiu}}, \bibinfo {author}
  {\bibfnamefont{J.-H.}\ \bibnamefont{Chung}}, \bibinfo {author}
  {\bibfnamefont{D.~G.}\ \bibnamefont{Nocera}},\ and\ \bibinfo {author}
  {\bibfnamefont{Y.~S.}\ \bibnamefont{Lee}},\ }%
  \bibfield{journal}{%
  \Doi{10.1103/PhysRevLett.98.107204}{\bibinfo {journal} {Phys. Rev. Lett.}}\
  }%
  \textbf{\bibinfo {volume} {98}},\ \bibinfo {pages} {107204} (\bibinfo {month}
  {Mar}\ \bibinfo {year} {2007})%
  \bibAnnoteFile{NoStop}{Helton:PRL:2007}%
\bibitem{Okamoto:PRL:2007}%
  \BibitemOpen
  \bibfield{author}{%
  \bibinfo {author} {\bibfnamefont{Y.}~\bibnamefont{Okamoto}}, \bibinfo
  {author} {\bibfnamefont{M.}~\bibnamefont{Nohara}}, \bibinfo {author}
  {\bibfnamefont{H.}~\bibnamefont{Aruga-Katori}},\ and\ \bibinfo {author}
  {\bibfnamefont{H.}~\bibnamefont{Takagi}},\ }%
  \bibfield{journal}{%
  \Doi{10.1103/PhysRevLett.99.137207}{\bibinfo {journal} {Phys. Rev. Lett.}}\
  }%
  \textbf{\bibinfo {volume} {99}},\ \bibinfo {pages} {137207} (\bibinfo {month}
  {Sep}\ \bibinfo {year} {2007})%
  \bibAnnoteFile{NoStop}{Okamoto:PRL:2007}%
\bibitem{Kandpal:PRL:2009}%
  \BibitemOpen
  \bibfield{author}{%
  \bibinfo {author} {\bibfnamefont{H.~C.}\ \bibnamefont{Kandpal}}, \bibinfo
  {author} {\bibfnamefont{I.}~\bibnamefont{Opahle}}, \bibinfo {author}
  {\bibfnamefont{Y.-Z.}\ \bibnamefont{Zhang}}, \bibinfo {author}
  {\bibfnamefont{H.~O.}\ \bibnamefont{Jeschke}},\ and\ \bibinfo {author}
  {\bibfnamefont{R.}~\bibnamefont{Valent\'\i{}}},\ }%
  \bibfield{journal}{%
  \Doi{10.1103/PhysRevLett.103.067004}{\bibinfo {journal} {Phys. Rev. Lett.}}\
  }%
  \textbf{\bibinfo {volume} {103}},\ \bibinfo {pages} {067004} (\bibinfo
  {month} {Aug}\ \bibinfo {year} {2009})%
  \bibAnnoteFile{NoStop}{Kandpal:PRL:2009}%
\bibitem{Nakamura:JPSJ:2009}%
  \BibitemOpen
  \bibfield{author}{%
  \bibinfo {author} {\bibfnamefont{K.}~\bibnamefont{Nakamura}}, \bibinfo
  {author} {\bibfnamefont{Y.}~\bibnamefont{Yoshimoto}}, \bibinfo {author}
  {\bibfnamefont{T.}~\bibnamefont{Kosugi}}, \bibinfo {author}
  {\bibfnamefont{R.}~\bibnamefont{Arita}},\ and\ \bibinfo {author}
  {\bibfnamefont{M.}~\bibnamefont{Imada}},\ }%
  \bibfield{journal}{%
  \Doi{10.1143/JPSJ.78.083710}{\bibinfo {journal} {Journal of the Physical
  Society of Japan}}\ }%
  \textbf{\bibinfo {volume} {78}},\ \bibinfo {pages} {083710} (\bibinfo {year}
  {2009})%
  \bibAnnoteFile{NoStop}{Nakamura:JPSJ:2009}%
\bibitem{Yoshioka:PRL:2009}%
  \BibitemOpen
  \bibfield{author}{%
  \bibinfo {author} {\bibfnamefont{T.}~\bibnamefont{Yoshioka}}, \bibinfo
  {author} {\bibfnamefont{A.}~\bibnamefont{Koga}},\ and\ \bibinfo {author}
  {\bibfnamefont{N.}~\bibnamefont{Kawakami}},\ }%
  \bibfield{journal}{%
  \Doi{10.1103/PhysRevLett.103.036401}{\bibinfo {journal} {Phys. Rev. Lett.}}\
  }%
  \textbf{\bibinfo {volume} {103}},\ \bibinfo {pages} {036401} (\bibinfo
  {month} {Jul}\ \bibinfo {year} {2009})%
  \bibAnnoteFile{NoStop}{Yoshioka:PRL:2009}%
\bibitem{Imada:Comment:2008}%
  \BibitemOpen
  \bibfield{author}{%
  \bibinfo {author} {\bibfnamefont{S.}~\bibnamefont{{Watanabe}}}, \bibinfo
  {author} {\bibfnamefont{T.}~\bibnamefont{{Mizusaki}}},\ and\ \bibinfo
  {author} {\bibfnamefont{M.}~\bibnamefont{{Imada}}}\ }%
  \Eprint{http://arxiv.org/abs/0811.3718}{arXiv:0811.3718}%
  \bibAnnoteFile{NoStop}{Imada:Comment:2008}%
\bibitem{Katsura:JSOP:1986}%
  \BibitemOpen
  \bibfield{author}{%
  \bibinfo {author} {\bibfnamefont{S.}~\bibnamefont{Katsura}}, \bibinfo
  {author} {\bibfnamefont{T.}~\bibnamefont{Ide}},\ and\ \bibinfo {author}
  {\bibfnamefont{T.}~\bibnamefont{Morita}},\ }%
  \bibfield{journal}{%
  \bibinfo {journal} {Journal of Statistical Physics}\ }%
  \textbf{\bibinfo {volume} {42}},\ \bibinfo {pages} {381} (\bibinfo {month}
  {02}\ \bibinfo {year} {1986})%
  \bibAnnoteFile{NoStop}{Katsura:JSOP:1986}%
\bibitem{Jolicoeur:PRB:1990}%
  \BibitemOpen
  \bibfield{author}{%
  \bibinfo {author} {\bibfnamefont{T.}~\bibnamefont{Jolicoeur}}, \bibinfo
  {author} {\bibfnamefont{E.}~\bibnamefont{Dagotto}}, \bibinfo {author}
  {\bibfnamefont{E.}~\bibnamefont{Gagliano}},\ and\ \bibinfo {author}
  {\bibfnamefont{S.}~\bibnamefont{Bacci}},\ }%
  \bibfield{journal}{%
  \Doi{10.1103/PhysRevB.42.4800}{\bibinfo {journal} {Phys. Rev. B}}\ }%
  \textbf{\bibinfo {volume} {42}},\ \bibinfo {pages} {4800} (\bibinfo {month}
  {Sep}\ \bibinfo {year} {1990})%
  \bibAnnoteFile{NoStop}{Jolicoeur:PRB:1990}%
\bibitem{Bernu:PRB:1994}%
  \BibitemOpen
  \bibfield{author}{%
  \bibinfo {author} {\bibfnamefont{B.}~\bibnamefont{Bernu}}, \bibinfo {author}
  {\bibfnamefont{P.}~\bibnamefont{Lecheminant}}, \bibinfo {author}
  {\bibfnamefont{C.}~\bibnamefont{Lhuillier}},\ and\ \bibinfo {author}
  {\bibfnamefont{L.}~\bibnamefont{Pierre}},\ }%
  \bibfield{journal}{%
  \Doi{10.1103/PhysRevB.50.10048}{\bibinfo {journal} {Phys. Rev. B}}\ }%
  \textbf{\bibinfo {volume} {50}},\ \bibinfo {pages} {10048} (\bibinfo {month}
  {Oct}\ \bibinfo {year} {1994})%
  \bibAnnoteFile{NoStop}{Bernu:PRB:1994}%
\bibitem{Capriotti:PRL:1999}%
  \BibitemOpen
  \bibfield{author}{%
  \bibinfo {author} {\bibfnamefont{L.}~\bibnamefont{Capriotti}}, \bibinfo
  {author} {\bibfnamefont{A.~E.}\ \bibnamefont{Trumper}},\ and\ \bibinfo
  {author} {\bibfnamefont{S.}~\bibnamefont{Sorella}},\ }%
  \bibfield{journal}{%
  \Doi{10.1103/PhysRevLett.82.3899}{\bibinfo {journal} {Phys. Rev. Lett.}}\ }%
  \textbf{\bibinfo {volume} {82}},\ \bibinfo {pages} {3899} (\bibinfo {month}
  {May}\ \bibinfo {year} {1999})%
  \bibAnnoteFile{NoStop}{Capriotti:PRL:1999}%
\bibitem{White:PRL:2007}%
  \BibitemOpen
  \bibfield{author}{%
  \bibinfo {author} {\bibfnamefont{S.~R.}\ \bibnamefont{White}}\ and\ \bibinfo
  {author} {\bibfnamefont{A.~L.}\ \bibnamefont{Chernyshev}},\ }%
  \bibfield{journal}{%
  \Doi{10.1103/PhysRevLett.99.127004}{\bibinfo {journal} {Phys. Rev. Lett.}}\
  }%
  \textbf{\bibinfo {volume} {99}},\ \bibinfo {pages} {127004} (\bibinfo {month}
  {Sep}\ \bibinfo {year} {2007})%
  \bibAnnoteFile{NoStop}{White:PRL:2007}%
\bibitem{Kalmeyer:PRB:1989}%
  \BibitemOpen
  \bibfield{author}{%
  \bibinfo {author} {\bibfnamefont{V.}~\bibnamefont{Kalmeyer}}\ and\ \bibinfo
  {author} {\bibfnamefont{R.~B.}\ \bibnamefont{Laughlin}},\ }%
  \bibfield{journal}{%
  \Doi{10.1103/PhysRevB.39.11879}{\bibinfo {journal} {Phys. Rev. B}}\ }%
  \textbf{\bibinfo {volume} {39}},\ \bibinfo {pages} {11879} (\bibinfo {month}
  {Jun}\ \bibinfo {year} {1989})%
  \bibAnnoteFile{NoStop}{Kalmeyer:PRB:1989}%
\bibitem{Morita:JPSC:2002}%
  \BibitemOpen
  \bibfield{author}{%
  \bibinfo {author} {\bibfnamefont{H.}~\bibnamefont{Morita}}, \bibinfo {author}
  {\bibfnamefont{S.}~\bibnamefont{Watanabe}},\ and\ \bibinfo {author}
  {\bibfnamefont{M.}~\bibnamefont{Imada}},\ }%
  \bibfield{journal}{%
  \Doi{10.1143/JPSJ.71.2109}{\bibinfo {journal} {Journal of the Physical
  Society of Japan}}\ }%
  \textbf{\bibinfo {volume} {71}},\ \bibinfo {pages} {2109} (\bibinfo {year}
  {2002})%
  \bibAnnoteFile{NoStop}{Morita:JPSC:2002}%
\bibitem{Sahebsara:PRL:2008}%
  \BibitemOpen
  \bibfield{author}{%
  \bibinfo {author} {\bibfnamefont{P.}~\bibnamefont{Sahebsara}}\ and\ \bibinfo
  {author} {\bibfnamefont{D.}~\bibnamefont{S\'en\'echal}},\ }%
  \bibfield{journal}{%
  \Doi{10.1103/PhysRevLett.100.136402}{\bibinfo {journal} {Phys. Rev. Lett.}}\
  }%
  \textbf{\bibinfo {volume} {100}},\ \bibinfo {pages} {136402} (\bibinfo
  {month} {Mar}\ \bibinfo {year} {2008})%
  \bibAnnoteFile{NoStop}{Sahebsara:PRL:2008}%
\bibitem{MetznerVollhardt:PRL:1989}%
  \BibitemOpen
  \bibfield{author}{%
  \bibinfo {author} {\bibfnamefont{W.}~\bibnamefont{Metzner}}\ and\ \bibinfo
  {author} {\bibfnamefont{D.}~\bibnamefont{Vollhardt}},\ }%
  \bibfield{journal}{%
  \Doi{10.1103/PhysRevLett.62.324}{\bibinfo {journal} {Phys. Rev. Lett.}}\ }%
  \textbf{\bibinfo {volume} {62}},\ \bibinfo {pages} {324} (\bibinfo {month}
  {Jan}\ \bibinfo {year} {1989})%
  \bibAnnoteFile{NoStop}{MetznerVollhardt:PRL:1989}%
\bibitem{Georges:RevModPhys:1996}%
  \BibitemOpen
  \bibfield{author}{%
  \bibinfo {author} {\bibfnamefont{A.}~\bibnamefont{Georges}}, \bibinfo
  {author} {\bibfnamefont{G.}~\bibnamefont{Kotliar}}, \bibinfo {author}
  {\bibfnamefont{W.}~\bibnamefont{Krauth}},\ and\ \bibinfo {author}
  {\bibfnamefont{M.~J.}\ \bibnamefont{Rozenberg}},\ }%
  \bibfield{journal}{%
  \Doi{10.1103/RevModPhys.68.13}{\bibinfo {journal} {Rev. Mod. Phys.}}\ }%
  \textbf{\bibinfo {volume} {68}},\ \bibinfo {pages} {13} (\bibinfo {month}
  {Jan}\ \bibinfo {year} {1996})%
  \bibAnnoteFile{NoStop}{Georges:RevModPhys:1996}%
\bibitem{Kotliar:RevModPhys:2006}%
  \BibitemOpen
  \bibfield{author}{%
  \bibinfo {author} {\bibfnamefont{G.}~\bibnamefont{Kotliar}}, \bibinfo
  {author} {\bibfnamefont{S.~Y.}\ \bibnamefont{Savrasov}}, \bibinfo {author}
  {\bibfnamefont{K.}~\bibnamefont{Haule}}, \bibinfo {author}
  {\bibfnamefont{V.~S.}\ \bibnamefont{Oudovenko}}, \bibinfo {author}
  {\bibfnamefont{O.}~\bibnamefont{Parcollet}},\ and\ \bibinfo {author}
  {\bibfnamefont{C.~A.}\ \bibnamefont{Marianetti}},\ }%
  \bibfield{journal}{%
  \Doi{10.1103/RevModPhys.78.865}{\bibinfo {journal} {Rev. Mod. Phys.}}\ }%
  \textbf{\bibinfo {volume} {78}},\ \bibinfo {pages} {865} (\bibinfo {month}
  {Aug}\ \bibinfo {year} {2006})%
  \bibAnnoteFile{NoStop}{Kotliar:RevModPhys:2006}%
\bibitem{Lu:arxiv:2007}%
  \BibitemOpen
  \bibfield{author}{%
  \bibinfo {author} {\bibfnamefont{F.}~\bibnamefont{Lu}}, \bibinfo {author}
  {\bibfnamefont{W.-H.}\ \bibnamefont{Wang}},\ and\ \bibinfo {author}
  {\bibfnamefont{L.-J.}\ \bibnamefont{Zou}}}%
   (\bibinfo {month} {09}\ \bibinfo {year} {2007}),\
  \Eprint{http://arxiv.org/abs/0709.4108v1}{0709.4108v1}%
  \bibAnnoteFile{NoStop}{Lu:arxiv:2007}%
\bibitem{Aryanpour:PRB:2006}%
  \BibitemOpen
  \bibfield{author}{%
  \bibinfo {author} {\bibfnamefont{K.}~\bibnamefont{Aryanpour}}, \bibinfo
  {author} {\bibfnamefont{W.~E.}\ \bibnamefont{Pickett}},\ and\ \bibinfo
  {author} {\bibfnamefont{R.~T.}\ \bibnamefont{Scalettar}},\ }%
  \bibfield{journal}{%
  \Doi{10.1103/PhysRevB.74.085117}{\bibinfo {journal} {Phys. Rev. B}}\ }%
  \textbf{\bibinfo {volume} {74}},\ \bibinfo {pages} {085117} (\bibinfo {month}
  {Aug}\ \bibinfo {year} {2006})%
  \bibAnnoteFile{NoStop}{Aryanpour:PRB:2006}%
\bibitem{Anderson:Book:1997}%
  \BibitemOpen
  \bibfield{author}{%
  \bibinfo {author} {\bibfnamefont{P.}~\bibnamefont{Anderson}},\ }%
  \emph{\bibinfo {title} {The Theory of Superconductivity in High-Tc
  Cuprates.}}\ (\bibinfo {publisher} {Princeton University Press},\ \bibinfo
  {year} {1997})%
  \bibAnnoteFile{NoStop}{Anderson:Book:1997}%
\bibitem{Licht:PRB:2000}%
  \BibitemOpen
  \bibfield{author}{%
  \bibinfo {author} {\bibfnamefont{A.~I.}\ \bibnamefont{Lichtenstein}}\ and\
  \bibinfo {author} {\bibfnamefont{M.~I.}\ \bibnamefont{Katsnelson}},\ }%
  \bibfield{journal}{%
  \Doi{10.1103/PhysRevB.62.R9283}{\bibinfo {journal} {Phys. Rev. B}}\ }%
  \textbf{\bibinfo {volume} {62}},\ \bibinfo {pages} {R9283} (\bibinfo {month}
  {Oct}\ \bibinfo {year} {2000})%
  \bibAnnoteFile{NoStop}{Licht:PRB:2000}%
\bibitem{Jarrell:RevModPhys:2005}%
  \BibitemOpen
  \bibfield{author}{%
  \bibinfo {author} {\bibfnamefont{T.}~\bibnamefont{Maier}}, \bibinfo {author}
  {\bibfnamefont{M.}~\bibnamefont{Jarrell}}, \bibinfo {author}
  {\bibfnamefont{T.}~\bibnamefont{Pruschke}},\ and\ \bibinfo {author}
  {\bibfnamefont{M.~H.}\ \bibnamefont{Hettler}},\ }%
  \bibfield{journal}{%
  \Doi{10.1103/RevModPhys.77.1027}{\bibinfo {journal} {Rev. Mod. Phys.}}\ }%
  \textbf{\bibinfo {volume} {77}},\ \bibinfo {pages} {1027} (\bibinfo {month}
  {Oct}\ \bibinfo {year} {2005})%
  \bibAnnoteFile{NoStop}{Jarrell:RevModPhys:2005}%
\bibitem{Hettler:PRB:1998}%
  \BibitemOpen
  \bibfield{author}{%
  \bibinfo {author} {\bibfnamefont{M.~H.}\ \bibnamefont{Hettler}}, \bibinfo
  {author} {\bibfnamefont{A.~N.}\ \bibnamefont{Tahvildar-Zadeh}}, \bibinfo
  {author} {\bibfnamefont{M.}~\bibnamefont{Jarrell}}, \bibinfo {author}
  {\bibfnamefont{T.}~\bibnamefont{Pruschke}},\ and\ \bibinfo {author}
  {\bibfnamefont{H.~R.}\ \bibnamefont{Krishnamurthy}},\ }%
  \bibfield{journal}{%
  \Doi{10.1103/PhysRevB.58.R7475}{\bibinfo {journal} {Phys. Rev. B}}\ }%
  \textbf{\bibinfo {volume} {58}},\ \bibinfo {pages} {R7475} (\bibinfo {month}
  {Sep}\ \bibinfo {year} {1998})%
  \bibAnnoteFile{NoStop}{Hettler:PRB:1998}%
\bibitem{Rubtsov:PRB:2008}%
  \BibitemOpen
  \bibfield{author}{%
  \bibinfo {author} {\bibfnamefont{A.~N.}\ \bibnamefont{Rubtsov}}, \bibinfo
  {author} {\bibfnamefont{M.~I.}\ \bibnamefont{Katsnelson}},\ and\ \bibinfo
  {author} {\bibfnamefont{A.~I.}\ \bibnamefont{Lichtenstein}},\ }%
  \bibfield{journal}{%
  \Doi{10.1103/PhysRevB.77.033101}{\bibinfo {journal} {Phys. Rev. B}}\ }%
  \textbf{\bibinfo {volume} {77}},\ \bibinfo {pages} {033101} (\bibinfo {month}
  {Jan}\ \bibinfo {year} {2008})%
  \bibAnnoteFile{NoStop}{Rubtsov:PRB:2008}%
\bibitem{Rubtsov:PRB:2009}%
  \BibitemOpen
  \bibfield{author}{%
  \bibinfo {author} {\bibfnamefont{A.~N.}\ \bibnamefont{Rubtsov}}, \bibinfo
  {author} {\bibfnamefont{M.~I.}\ \bibnamefont{Katsnelson}}, \bibinfo {author}
  {\bibfnamefont{A.~I.}\ \bibnamefont{Lichtenstein}},\ and\ \bibinfo {author}
  {\bibfnamefont{A.}~\bibnamefont{Georges}},\ }%
  \bibfield{journal}{%
  \Doi{10.1103/PhysRevB.79.045133}{\bibinfo {journal} {Phys. Rev. B}}\ }%
  \textbf{\bibinfo {volume} {79}},\ \bibinfo {pages} {045133} (\bibinfo {month}
  {Jan}\ \bibinfo {year} {2009})%
  \bibAnnoteFile{NoStop}{Rubtsov:PRB:2009}%
\bibitem{Liebsch:PRB:2009}%
  \BibitemOpen
  \bibfield{author}{%
  \bibinfo {author} {\bibfnamefont{A.}~\bibnamefont{Liebsch}}, \bibinfo
  {author} {\bibfnamefont{H.}~\bibnamefont{Ishida}},\ and\ \bibinfo {author}
  {\bibfnamefont{J.}~\bibnamefont{Merino}},\ }%
  \bibfield{journal}{%
  \Doi{10.1103/PhysRevB.79.195108}{\bibinfo {journal} {Phys. Rev. B}}\ }%
  \textbf{\bibinfo {volume} {79}},\ \bibinfo {pages} {195108} (\bibinfo {month}
  {May}\ \bibinfo {year} {2009})%
  \bibAnnoteFile{NoStop}{Liebsch:PRB:2009}%
\bibitem{LeeMonien:PRB:2008}%
  \BibitemOpen
  \bibfield{author}{%
  \bibinfo {author} {\bibfnamefont{H.}~\bibnamefont{Lee}}, \bibinfo {author}
  {\bibfnamefont{G.}~\bibnamefont{Li}},\ and\ \bibinfo {author}
  {\bibfnamefont{H.}~\bibnamefont{Monien}},\ }%
  \bibfield{journal}{%
  \Doi{10.1103/PhysRevB.78.205117}{\bibinfo {journal} {Phys. Rev. B}}\ }%
  \textbf{\bibinfo {volume} {78}},\ \bibinfo {pages} {205117} (\bibinfo {month}
  {Nov}\ \bibinfo {year} {2008})%
  \bibAnnoteFile{NoStop}{LeeMonien:PRB:2008}%
\bibitem{Kyung:PRB:2007}%
  \BibitemOpen
  \bibfield{author}{%
  \bibinfo {author} {\bibfnamefont{B.}~\bibnamefont{Kyung}},\ }%
  \bibfield{journal}{%
  \Doi{10.1103/PhysRevB.75.033102}{\bibinfo {journal} {Phys. Rev. B}}\ }%
  \textbf{\bibinfo {volume} {75}},\ \bibinfo {pages} {033102} (\bibinfo {month}
  {Jan}\ \bibinfo {year} {2007})%
  \bibAnnoteFile{NoStop}{Kyung:PRB:2007}%
\bibitem{Rubtsov:PRB:2005}%
  \BibitemOpen
  \bibfield{author}{%
  \bibinfo {author} {\bibfnamefont{A.~N.}\ \bibnamefont{Rubtsov}}, \bibinfo
  {author} {\bibfnamefont{V.~V.}\ \bibnamefont{Savkin}},\ and\ \bibinfo
  {author} {\bibfnamefont{A.~I.}\ \bibnamefont{Lichtenstein}},\ }%
  \bibfield{journal}{%
  \Doi{10.1103/PhysRevB.72.035122}{\bibinfo {journal} {Phys. Rev. B}}\ }%
  \textbf{\bibinfo {volume} {72}},\ \bibinfo {pages} {035122} (\bibinfo {month}
  {Jul}\ \bibinfo {year} {2005})%
  \bibAnnoteFile{NoStop}{Rubtsov:PRB:2005}%
\bibitem{Hafermann:2007:JETPLett}%
  \BibitemOpen
  \bibfield{author}{%
  \bibinfo {author} {\bibfnamefont{H.}~\bibnamefont{{Hafermann}}}, \bibinfo
  {author} {\bibfnamefont{S.}~\bibnamefont{{Brener}}}, \bibinfo {author}
  {\bibfnamefont{A.~N.}\ \bibnamefont{{Rubtsov}}}, \bibinfo {author}
  {\bibfnamefont{M.~I.}\ \bibnamefont{{Katsnelson}}},\ and\ \bibinfo {author}
  {\bibfnamefont{A.~I.}\ \bibnamefont{{Lichtenstein}}},\ }%
  \bibfield{journal}{%
  \Doi{10.1134/S0021364007220134}{\bibinfo {journal} {JETP Lett.}}\ }%
  \textbf{\bibinfo {volume} {86}},\ \bibinfo {pages} {677} (\bibinfo {month}
  {Dec.}\ \bibinfo {year} {2007}),\
  \Eprint{http://arxiv.org/abs/0707.4022}{arXiv:0707.4022}%
  \bibAnnoteFile{NoStop}{Hafermann:2007:JETPLett}%
\bibitem{Kyung:PRL:2006}%
  \BibitemOpen
  \bibfield{author}{%
  \bibinfo {author} {\bibfnamefont{B.}~\bibnamefont{Kyung}}\ and\ \bibinfo
  {author} {\bibfnamefont{A.-M.~S.}\ \bibnamefont{Tremblay}},\ }%
  \bibfield{journal}{%
  \Doi{10.1103/PhysRevLett.97.046402}{\bibinfo {journal} {Phys. Rev. Lett.}}\
  }%
  \textbf{\bibinfo {volume} {97}},\ \bibinfo {pages} {046402} (\bibinfo {month}
  {Jul}\ \bibinfo {year} {2006})%
  \bibAnnoteFile{NoStop}{Kyung:PRL:2006}%
\bibitem{Hafermann:PRL:2009}%
  \BibitemOpen
  \bibfield{author}{%
  \bibinfo {author} {\bibfnamefont{H.}~\bibnamefont{Hafermann}}, \bibinfo
  {author} {\bibfnamefont{G.}~\bibnamefont{Li}}, \bibinfo {author}
  {\bibfnamefont{A.~N.}\ \bibnamefont{Rubtsov}}, \bibinfo {author}
  {\bibfnamefont{M.~I.}\ \bibnamefont{Katsnelson}}, \bibinfo {author}
  {\bibfnamefont{A.~I.}\ \bibnamefont{Lichtenstein}},\ and\ \bibinfo {author}
  {\bibfnamefont{H.}~\bibnamefont{Monien}},\ }%
  \bibfield{journal}{%
  \Doi{10.1103/PhysRevLett.102.206401}{\bibinfo {journal} {Phys. Rev. Lett.}}\
  }%
  \textbf{\bibinfo {volume} {102}},\ \bibinfo {pages} {206401} (\bibinfo
  {month} {May}\ \bibinfo {year} {2009})%
  \bibAnnoteFile{NoStop}{Hafermann:PRL:2009}%
\end{thebibliography}%
\end{document}